\documentclass[pss,fleqn,embeddedheads]{w-art}
\usepackage{times}
\usepackage{w-thm}
\usepackage[]{graphicx}
\def\bc{\begin{center}}
\def\ec{\end{center}}

\def\beq{\begin{equation}}
\def\eeq{\end{equation}}

\def\d{\downarrow}
\def\u{\uparrow}

\def\bj{{\bf j}}

\begin{document}

\DOIsuffix{theDOIsuffix}
\Volume{XX}
\Issue{1}
\Copyrightissue{01}
\Month{01}
\Year{2004}
\pagespan{1}{}
\title[Orbital Ordering]{Electron-phonon interaction in correlated 
electronic systems: polarons and the formation of orbital ordering}
\author[Schneider]{D. Schneider\footnote{Corresponding
     author: e-mail: {\sf daniela.schneider@physik.uni-augsburg.de}, 
Phone: +49\,821\,598\,3221,\\ Fax:
     +49\,821\,598\,3262}}
\address{Institut f\"ur Physik, Universit\"at Augsburg, 
Universit\"atsstrasse 1, 86135 Augsburg, Germany}
\author[H\"ock]{K.-H. H\"ock}
\author[Ziegler]{K. Ziegler}
\begin{abstract}
The properties of a dilute electron gas, coupled to the lattice degrees of freedom, are studied and compared with the properties of an electron gas at half-filling, where spinless fermions with two orbitals per lattice site are considered. The 
simplest model which includes both the local electron-lattice interaction of the Jahn-Teller type and the electronic correlations is the $E\otimes\beta$-Jahn-Teller-Hubbard
model. We analyze the formation and 
stability of Jahn-Teller polarons and bipolarons, respectively. Our approach is 
based on a hopping expansion in the strong-coupling regime. The results are compared 
with recently published findings for the Hubbard-Holstein model 
\cite{sawatzky,bonca}.
The special case of the Jahn-Teller-Hubbard model at half-filling is mapped on 
a spin-1/2 Heisenberg model with phonon-dependent coupling constants. This 
has been derived within a projection formalism that provides a continued-fraction 
representation of the Green's function. 
We study the exact solution for two and three particles 
and compare it with the effective theory on the infinite lattice with one 
particle per site.
\end{abstract}

\maketitle

\section{The Model}

The Hamiltonian for fermions with spin $\sigma$ and pseudospin $\gamma=\theta ,\epsilon$,
coupled to phonons, is given by $H=H_t+H_0$,
where $H_t$ is a nearest neighbor tunneling term for the fermions
\[
H_t=-t\sum_{<\bj,\bj'>}\sum_{\sigma=\u,\d}\sum_{\gamma=\theta,\epsilon}
c_{\bj\gamma\sigma}^\dagger c_{\bj'\gamma\sigma}+h.c.
\]
and $H_0$ is local, containing a (Hubbard) interaction and a
term for dispersionless phonons of energy $\omega_0$ with $H_0=\sum_{\bj}H_{0\bj}$
and
\[
H_{0\bj}=\omega_0 b_\bj^\dagger b_\bj
+g(b_\bj^\dagger+b_\bj)\sum_\sigma(n_{\bj\theta\sigma}\pm n_{\bj\epsilon\sigma})
+\sum_\gamma Un_{\bj\gamma\u}n_{\bj\gamma\d}+U_O (n_{\bj\theta\u}+n_{\bj\theta\d})
(n_{\bj\epsilon\u}+n_{\bj\epsilon\d}),
\]
where the plus (minus) sign refers to Holstein ($E\otimes\beta$ Jahn-Teller)
electron-phonon coupling \cite{englman}. $H_{0\bj}$ is diagonalized by a Lang-Firsov 
transformation \cite{langfirsov} and has energies
\beq
E_{0\bj}
=\omega_0 n_\bj-E_{p}
\Big[\sum_\sigma (n_{\bj\theta\sigma}\pm n_{\bj\epsilon\sigma})\Big]^2
+\sum_\gamma Un_{\bj\gamma\u}n_{\bj\gamma\d}+U_O (n_{\bj\theta\u}+n_{\bj\theta\d})
(n_{\bj\epsilon\u}+n_{\bj\epsilon\d})
\label{energy0}
\eeq
if there are ${n_\bj}(\ge0)$ phonons and $n_{\bj\gamma\sigma}(=0,1)$ electrons
with (pseudo)spin $\sigma$ ($\gamma$) at site ${\bf j}$. Each electron has 
an energy gain $E_{p}=g^2/\omega_{0}$. 
The regime $U<2E_{p}$ has an attractive interaction, leading to an
on-site bipolaron.

\section{The Bipolaron Problem}

To study the binding properties of polarons we consider two electrons on a lattice with 
two orbitals ($\theta$ and $\epsilon$) per lattice site. In a strong coupling approach
we study the similarities and differences of the non-degenerate Holstein-Hubbard ($n_\epsilon$ or $n_\theta=0$) (HH) and the
$E\otimes\beta$-Jahn-Teller-Hubbard model (JTH). We restrict ourselves to the case of repulsive on-site interaction
($U>2E_{p}$). Therefore the ground state of $H_0$ has single occupied sites. If both electrons are in the same orbital, the situation is the same as in the HH 
\cite{sawatzky,bonca} due to the orbital conserving hopping. We show in the following, however, that the 
situation for electrons in the $\theta$ and the $\epsilon$ orbitals is different.
This is a consequence of the fact that we have a different distortion configuration associated with different orbitals.

For a hypercubic lattice a hopping expansion 
leads to an exponentially reduced hopping rate $te^{-E_{p}/\omega_{0}}$ in first order. The results in
second order for each possible hopping process can be written in general as
\beq
-\;\frac{t^2}{\omega_{0}}\;e^{-\frac{2E_{p}}{\omega_{0}}}\sum_{n=1}^{\infty}\frac{(-b)^{n}}{n!\;n}
\quad\mbox{and}\quad
-\;\frac{t^2}{\omega_{0}}\;e^{-\frac{2E_{p}}{\omega_{0}}}\sum_{n=0}^{\infty}\frac{(-b)^{n}}{n!\;
(n+a)}=-\;\frac{t^2}{\omega_{0}}\;e^{-\frac{2E_{p}}{\omega_{0}}}\;{\tilde\gamma}(a,b)\ \ .
\eeq
The first series is the result of processes with empty and singly-occupied sites as 
excited states, while the second series takes doubly-occupied states into account. ${\tilde\gamma}(a,b)$ 
is related to the incomplete Gamma function \cite{abramowitz}. For electrons
in the same orbital $a$ is equal to $(U-2E_{p})/\omega_{0}$ and for different orbital
occupation  $a$ is given as $(U_O +2E_{p})/\omega_{0}$. Values of $b$ are
$\pm E_{p}/\omega_{0}$ and  $\pm 2E_{p}/\omega_{0}$. The positive sign corresponds only to processes that describe hopping in different orbitals.
One process takes place in the $\theta$, the other in the $\epsilon$ orbital. Hence they only appear in the JTH but not in the HH.
The terms related to the processes explained above are related to processes that change the orbital occupancy
at least at one site. This means, for instance, that the initial state has an electron in orbital $\theta$ and the final state an electron in orbital $\epsilon$ at one site. Only the electrons at nearest neighbor
sites are available for these processes. If both hopping processes take
place in the same orbital the minus sign appears.
These are the only contributions in the HH. They also exist in the JTH
if either both electrons
are in the same orbital or if only one of the two electrons moves. The factor $2$ arises if
only two sites are involved in the process. 
While the series with $b>0$ reduces the exponential prefactor even further, the series with $b<0$ cancel
it in part or, in the case of $-2E_{p}/\omega_{0}$, even completely. This means that hopping processes that
do not change the lattice configuration, i. e. the orbital occupancy, are not suppressed exponentially in the strong coupling regime. Processes with exchange in the same orbital
are included 
in this case but not those in different orbitals. The latter change the orbital occupancy
and therefore the lattice configuration.
In a strong coupling approach we neglect the exponentially reduced terms. For nearest-neighbor
electrons in the same orbital we have the possibility of an exchange, which eventually leads to 
a nearest-neighbor singlet and triplet configuration as in the HH. The singlet, a nearest-neighbor bipolaron,
is the ground state if $U<4E_{p}$. The corresponding binding energy 
does not dependent on the coordination number of the lattice. This is an artefact of our approximation,
and it is $\Delta=t^2/E_{p}-4t^2/U$. On the other hand, the exchange of electrons in different orbitals
vanishes in the strong coupling regime. The lowest energy state of electrons
in different orbitals is that of unbound polarons, and it is equal to that of unbound polarons in
the same orbital.
The nearest neighbor states are higher in energy with a finite energy shift with respect
to the unbound states.

\section{Jahn-Teller Effect for Systems with One Particle per Site}

For simplicity, the spin of the electron
is ignored now. Formally, the orbital has a similar meaning as
the spin in the sense that it appears as an additional degree of 
freedom. This gives us orbital-depending effects in tunneling,
very similar to spin-dependent effects \cite{loss04}. In this case the ground state for 
sufficiently small hopping rate $t$ compared with $E_{p}$, $\omega_0$ and  $U_O$ is a singly occupied lattice without
phonons.  

In the following we shall consider the Green's function $G(z)=(z-H)^{-1}$ and its projection to the subspace with one fermion per
lattice site and no phonon $P_0$. The Green's function satisfies the
operator identity \cite{ziegler03}
\beq
P_0(z-H)^{-1}P_0=
(P_0(z-H)P_0 - P_0H P_1 (z-H)_1^{-1}P_1 HP_0)_0^{-1}
\equiv (z-P_0H_0P_0 - H_{eff})_0^{-1},
\label{projected}
\eeq
where $P_1$ projects to the complement: $P_1=1-P_0$. $(...)_j^{-1}$ is
the inverse on the projected Hilbert space. Since the 
Hamiltonian $H$ preserves the total number of fermions in the system, the
matrix of $H$ in fermion-number representation is given
by diagonal blocks. Consequently, the projected 
matrix $H_{eff}=P_0H P_1 (z-H)_1^{-1}P_1 HP_0$
acts only on the Hilbert space with a total fermion number
equal to the number of lattice sites. In other words, all
virtual processes in $H_{eff}$ are creation and annihilation
processes of pairs of doubly occupied and empty sites. 
Equation (\ref{projected}) is a recursion relation on the projected
Hilbert spaces and can be iterated. It provides a 
continued-fraction representation \cite{ziegler03}. 
A truncation after the first iteration by replacing $H\to H_0$ in
$(z-H)_1^{-1}$ leads to the approximation of $H_{eff}$ by 
the XXZ Heisenberg model
\beq
H_{eff}\approx\frac{t^2}{2}\sum_{<{\bf j},{\bf j}'>}\Big[
a_{\u\u}(S_{\bf j}^zS_{{\bf j}'}^z-1)+
a_{\u\d}(S_{\bf j}^xS_{{\bf j}'}^x+S_{\bf j}^yS_{{\bf j}'}^y)
\Big]
\eeq
with coupling constants for a lattice with $N$ sites
\begin{eqnarray}
a_{\u\u}(z)&=&-4\frac{e^{-2E_{p}/\omega_0}}{\omega_0}
{\tilde\gamma}\left(\frac{U_O-z-(N-2)E_{p}}{\omega_0},-\frac{2E_{p}}{\omega_0}\right)\quad\mbox{and}\nonumber\\
a_{\u\d}(z)&=&-4\frac{e^{-2E_{p}/\omega_0}}{\omega_0}
{\tilde\gamma}\left(\frac{U_O-z-(N-2)E_{p}}{\omega_0},\frac{2E_{p}}{\omega_0}\right).
\end{eqnarray}
For weak and strong electron-phonon coupling we obtain
the same behavior for $a_{\u\u}$:
\beq
a_{\u\u}\sim\frac{4}{z+(N-2)E_{p}-U_O}
\eeq
but a different behavior for $a_{\u\d}$, where 
\beq
a_{\u\d}\sim a_{\u\u}\ \ \ \ (g/\omega_0\ll 1), \ \ \ 
a_{\u\d}\sim 0\ \ \ \ (g/\omega_0\gg 1).
\eeq
Thus the weak-coupling limit of $H_{eff}$ gives an isotropic 
Heisenberg model, whereas the strong-coupling 
limit leads to an Ising model.

\subsection{\bf Three-atomic Molecules}

As a simple example we study the effect of the geometry on
properties in the case of a molecule with three atoms ($N=3$),
either in a stretched (SM) or triangular (TM) configuration.
The Hilbert space of the full Hamiltonian has infinite dimensions,
whereas the Hilbert space of the Hamiltonian of the projected 
Green's function with one electron per site and no phonon
has only dimension $d=8$. The problem of diagonalizing the effective
Hamiltonian is further reduced by the global spin-flip symmetry to $d=4$.
However, the eigenvalues of $H_{eff}$ are complicated functions
of $z$ due to the virtual hopping processes which include the 
creation of an arbitrary number of phonons. The projected Green's
function of Eq. (\ref{projected}) can be expressed in terms of 
eigenstates and eigenvalues of the effective Hamiltonian $H_{eff}$ as $\langle E_j|(z-H)^{-1}|E_j\rangle=1/(z-E_j(z)).$ The energy functions $E_j(z)$ for the stretched and for the triangular 
molecule are listed in Table \ref{energies}. The geometric degeneracy of the TM
gives a doubly degenerate ground state for $E_0$. Therefore, the energy 
$E_2$ is absent in this case.
\begin{table}[t]
\begin{tabular}{ccc}
   & TM & SM \\[0.2cm]
$E_0$ & $-3E_{p}+t^2(a_{\u\u}+a_{\u\d}/2)$
 & $-3E_{p}-t^2\lambda_0/2$ \\
$E_1$ & $-3E_{p}+t^2(a_{\u\u}-a_{\u\d})$ 
 & $-3E_{p}+t^2a_{\u\u}/2$ \\
$E_2$ & -- & $-3E_{p}-t^2\lambda_2/2$ \\
$E_3$ & $-3E_{p}$ & $-3E_{p}$ \\[0.5cm]
\end{tabular}
\hspace*{1cm}
\begin{tabular}{ccccc}
      & TM/ & TM/ & SM/ & SM/ \\
      & WC  & SC  & WC  & SC \\[0.2cm]
$E_0$ & $3$   &  $2$  &  $3$  & $2$  \\
$E_1$ & $0$   &  $2$  &  $1$  & $1$  \\
$E_2$ & $-$   &  $-$  &  $0$  & $1$  \\
$E_3$ & $0$   &  $0$  &  $0$  & $0$  \\
\end{tabular}
\parbox[t]{0.7\textwidth}{
\caption[smallcaption]{
Energies for the TM and the SM with\\
$\lambda_0=-\frac{3}{2}a_{\u\u}-\frac{1}{2}\sqrt{a_{\u\u}^2+8a_{\u\d}^2}$ and 
$\lambda_2=-\frac{3}{2}a_{\u\u}+\frac{1}{2}\sqrt{a_{\u\u}^2+8a_{\u\d}^2}$.}
\label{energies}}
\parbox[t]{0.25\textwidth}{
\caption[smallcaption]{Values of the parameter $\lambda$ in Eq. (\ref{gf2}).}
\label{lambda}}
\end{table}

\subsection{\bf Discussion of the pole structure}

The energy levels of the molecules are poles of the projected 
Green's functions $1/(z-E_j)$, i.e. they are solutions of the
equation
$
z=E_j(z)
$.
The parameter $z$ appears in $a_{\u\u}$ and $a_{\u\d}$ 
(and therefore in $E_j$) only in the combination $z'=z+E_{p}$.
In general, we expect a complex solution $z'=x+iy$ for $z=E_j$. However,
it turns out from the properties of the incomplete Gamma function that the
imaginary part is always $y=0$ and that the real part satisfies an 
equation of the form
\beq
x=\frac{t^2}{2}\sum_{m}\frac{c_{m}}{x-\alpha_{m}}\ \ \ 
(c_{m}\ge0,\ \ \ 
\alpha_{m}=\omega_0m-2E_{p}+U_O).
\eeq
Consequently, there are solutions $x_{m}$ with $\alpha_{m}<x_{m}<\alpha_{m}+\omega_0$.

\subsubsection{\bf Pole structure in the asymptotics}

In weak coupling (WC) as well as strong coupling (SC), the Green's 
functions
have only two poles for each $E_j$. These poles are solutions of 
a quadratic equation. 
To study all elements of the Green's functions in one case,
the parameter $\lambda$ is introduced such that the Green's function
reads
\beq
\langle E_j|(z-H)^{-1}|E_j\rangle
=\frac{1}{z'-\frac{2t^2\lambda}{z'-2E_{p}-U_O}},
\ \ (z'=z+3E_{p})
\label{gf2}
\eeq
with the values of the parameter $\lambda$ given in Table \ref{lambda}. The poles in Eq. (\ref{gf2}) are
\beq
z'_{1/2}=E_{p}+U_{O}/2\pm\sqrt{(E_{p}+U_{O}/2)^2+2t^2\lambda}.
\eeq
The ground state is related to $min\{z'_1,z'_2\}$, i.e., it is $z'_2$.
Moreover, $\lambda$ must be maximal. Thus $E_0$ is the ground state, except
for TM/SC, where the ground state has an additional degeneracy due to 
$E_0=E_1$. 

\section{\bf Conclusions}

It is argued in \cite{bonca} that in the strong coupling 
regime the main source of bipolaron formation is the non-exponential 
off-diagonal matrix element in second order related to the exchange of 
neighboring electrons. In the $E\otimes\beta$ case we found exponential 
decay for this exchange in the situation of different orbital occupancy. 
Therefore the unbound state is preferred. For one particle per site we showed 
that $H_{eff}$ of the projected Green's function yields an XXZ Heisenberg 
model. Furthermore it should be noted that the corresponding Holstein-Hubbard 
model leads to an isotropic Heisenberg term. This was also discussed for the bipolaron problem \cite{sawatzky,bonca}.  As an application we have discussed 
the spectral properties for stretched and triangular molecular configurations 
of three atoms.

\vskip0.5cm

\begin{acknowledgement}
This work was supported by the Deutsche Forschungsgemeinschaft through
Sonderforschungsbereich 484.
\end{acknowledgement}

\end{document}